\documentclass[aps,prb,amsmath,twocolumn]{revtex4}
\usepackage{graphicx}
\begin{document}

\title{Persistent currents in nanorings and quantum decoherence by Coulomb
  interaction}
\author{Andrew G. Semenov$^{1}$
and Andrei D. Zaikin$^{2,1}$
}
\address{$^1$I.E.Tamm Department of Theoretical Physics, P.N.Lebedev
Physics Institute, 119991 Moscow, Russia\\
$^2$Forschungszentrum Karlsruhe, Institut f\"ur Nanotechnologie,
76021 Karlsruhe, Germany
}

\begin{abstract}
Employing instanton technique we evaluate equilibrium persistent
current (PC) produced by a quantum particle moving in a periodic
potential on a ring and interacting with a dissipative environment
formed by diffusive electron gas. The model allows for detailed
non-perturbative analysis of interaction effects and -- depending
on the system parameters -- yields a rich structure of different
regimes. We demonstrate that at low temperatures PC is
exponentially suppressed at sufficiently large ring perimeters
$2\pi R> L_{\varphi}$ where the dephasing length $L_{\varphi}$ is
set by interactions and does not depend on temperature. This
behavior represents a clear example of quantum decoherence by
electron-electron interactions at $T\to 0$.
\end{abstract}

\maketitle

\section{Introduction}

Electron decoherence is one of the key ingredients of the
many-body ground state in the presence of disorder and
electron-electron interactions. The existing non-perturbative
theory of this phenomenon at low temperatures in realistic
disordered conductors \cite{GZ1,GZ2,SGZ} is rather complicated, to
a large extent because of the necessity to properly account for
Fermi statistics for interacting electrons. At the same time the
main physical reason for electron dephasing appears obvious
already without unnecessary technical details: It is the electron
interaction with the fluctuating quantum electromagnetic field
produced by other electrons moving in a disordered potential.

In order to be able to quantitatively describe and understand the
latter effect Guinea \cite{Paco} suggested a model which mimics
all essential features of the ``real'' problem of interacting
electrons in a disordered conductor except for the Pauli exclusion
principle. This model describes a quantum particle moving on a
ring with radius $R$ and interacting with quantum dissipative
environment. For a system in thermodynamic equilibrium quantum
decoherence manifests itself as effective suppression of
off-diagonal density matrix elements beyond a certain length
$L_{\varphi}$. Provided there exists nonzero electron dephasing
due to its interaction with quantum environment at $T\to 0$, this
dephasing length $L_{\varphi}$ should stay finite down to zero
temperature. Hence, all effects sensitive to quantum coherence,
such as, e.g., persistent currents (PC) and Aharonov-Bohm (AB)
oscillations in mesoscopic rings, should be suppressed by
interactions as soon as the ring perimeter $2\pi R$ exceeds
$L_{\varphi}$.

A great deal of information can be obtained by modelling the
environment by a bath of Caldeira-Leggett (CL) oscillators. In
this case it was demonstrated \cite{Buttiker} that PC is reduced
by interactions {\it in the ground state} implying suppression of
quantum coherence exactly at $T=0$. For the same CL environment
Guinea \cite{Paco} found that AB oscillations for a quantum
particle on a ring are suppressed by the factor $\sim \exp
(-(R/L_{\varphi})^2)$, where the length $L_{\varphi}$ is set by
interactions and remains finite down to $T=0$. A similar result
was also obtained earlier from the real-time analysis \cite{GZ98}.
Furthermore, the problem \cite{Paco,GZ98} is exactly equivalent to
that of Coulomb blockade in the so-called single electron box
where exponential reduction of the effective charging energy at
large conductances \cite{pa91,HSZ} is presently considered as a
well established result. Thus, it is now widely accepted that PC
for a quantum particle on a ring is exponentially reduced at large
ring perimeters down to $T=0$ due to strong dephasing produced by
interaction between the particle and the CL bath.

It appears that presently no such consensus exists for another
important model of the environment \cite{Paco} formed by a
diffusive electron gas. Renormalization group arguments developed
for this model \cite{Paco} suggest very weak power law $\sim
R^{-\chi }$ suppression of AB oscillations at $T \to 0$, where the
factor $\chi \ll 1$ is set by interactions. On the contrary, the
combination of semiclassics, instanton technique and quantum Monte
Carlo (MC) analysis \cite{GHZ} yields much stronger suppression of
quantum coherence, namely exponential suppression $\sim \exp
(-R/L_{\varphi})$ (with temperature independent $L_{\varphi}$) at
not too low $T$ and power law suppression $\sim R^{-\chi}$ with
$\chi \approx 1.8$ at $T \to 0$ and for $2\pi R \gtrsim
L_{\varphi}$.

More recently this problem was reconsidered by means of
variational approach \cite{HlD}, perturbation theory \cite{CH} and
MC simulations \cite{KH}. In contrast to Ref. \onlinecite{GHZ},
either no \cite{HlD,CH} or very weak \cite{KH} $R$-dependent
suppression of PC was found.
Note, however, that the variational calculation \cite{HlD} reduces the PC problem to that of  mass renormalization in the $m=0$ topological sector while, as we show here, PC is determined by other topological sectors ($m\neq 0$) that are distinct from the $m=0$ sector, unlike the variational result in Ref. \onlinecite{HlD}.
Perturbative in the interaction calculations \cite{CH} can also
miss the correct behavior of PC at not too small $R$ as it was
already demonstrated in Ref. \onlinecite{GHZ} for the problem
under consideration and was also discussed elsewhere in a broader
context \cite{GZ1,GZ2}. More arguments along the same lines will
be presented below in this paper.

As far as numerical MC results are concerned, the authors
\cite{KH} ascribed the difference between their conclusions and
those of the previous work \cite{GHZ} to insufficient Trotter
number values employed in the MC analysis \cite{GHZ}. While this
particular issue definitely requires further analysis, it is worth
pointing out that the MC data \cite{KH} cover only the
perturbative regime $R < L_{\varphi}$ where rather weak
suppression of PC was found in Ref. \onlinecite{GHZ} as well, cf.
Fig. 3 in that paper. Hence, MC results \cite{KH} for small $R$ do
not appear conclusive for the most interesting non-perturbative
regime $R \gg L_{\varphi}$ where strong (though weaker than
exponential) interaction-induced suppression of PC was predicted
\cite{GHZ} down to $T \to 0$.

Leaving detailed discussion of MC procedure and results to the
future, here we note that -- despite significant theoretical
activity in the field -- currently no well-controlled
non-perturbative approach is available which would allow to
analytically study PC for the ``particle + dirty electron gas''
model \cite{Paco} in the low temperature limit and at sufficiently
large values of $R$. Given both fundamental importance of the
problem and remaining controversies in the literature, it is
highly desirable to formulate such an approach which would
unambiguously resolve the whole issue. A step in this direction is
undertaken in the present work.

The structure of our paper is as follows. In Sec. 2 we define our
model which is essentially identical to that pioneered by Guinea
\cite{Paco} except we additionally introduce a periodic potential
for the particle on a ring. In Sec. 3 we formulate our main
formalism and evaluate persistent current in the ring without
interactions by means of the standard instanton technique. This
approach is conveniently generalized to the interacting regime in
Sec. 4 where we analyze suppression of both quantum coherence and
PC by Coulomb interaction at sufficiently low temperatures down to
$T \to 0$ and arbitrarily large ring perimeters. Discussion of our
main observations and conclusions is presented in Sec. 5. Some
technical details of our calculation are specified in Appendices A
and B.

\section{The model and effective action}

\begin{figure}[t]
\includegraphics[width=3.375in]{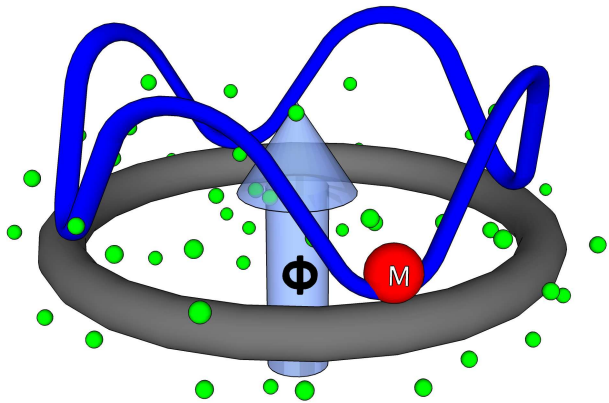}
\caption{\label{f1} (Color online) The system under consideration:
A particle on a ring in the presence of a periodic potential. The
ring is pierced by the magnetic flux and the particle interacts
with an effective environment formed by a dirty electron gas.}
\end{figure}

We will consider a quantum particle with mass $M$ and electric
charge $e$ on a ring with radius $R$ threaded by external magnetic
flux $\Phi_x$, see Fig. \ref{f1}. As before \cite{Paco,GHZ}, it will be
convenient to describe the particle position by a vector
$\bbox{R}=(R\cos \theta ,R\sin \theta )$ and consider the angle
$\theta$ as a quantum variable. In contrast to Refs.
\onlinecite{Paco,GHZ,HlD,CH,KH} where the quantum particle on a
ring was described only by its kinetic energy (i.e. no potential
energy was included into consideration), in this paper we will
assume that the particle moves in a periodic potential which --
just for the sake of definiteness -- is chosen in the form
$U(\theta )= U_0(1-\cos (\kappa \theta ))$. Here $\kappa$ is the
total number of periods of the potential $U(\theta )$ which the
particle should pass before it makes one full circle on the ring.
Accordingly, a non-interacting particle on a ring is described by
the Hamiltonian
\begin{equation}
\hat H_0= \frac{E_C(\hat \Phi +\Phi_x)^2}{\Phi_0^2} + U_0(1-\cos
(\kappa \theta )),
\label{H0}
\end{equation} where $\hat
\Phi=-i\Phi_0\partial /\partial \theta$ is the magnetic flux
operator, $E_C=1/(2MR^2)$ and $\Phi_0=2\pi c/e$ is the flux
quantum (here and below we set the Planck's constant equal to
unity $\hbar =1$).

Now let us include the interaction between the particle on a ring
and an effective dissipative environment. Specifically, we will
assume that the ring is embedded in the environment formed by the
so-called ``dirty electron gas'' \cite{Paco}. The total
Hamiltonian for our system reads
\begin{equation}
\hat H=\hat H_0+\hat H_{\rm el}+\hat H_{\rm int},
\end{equation}
where $\hat H_{\rm el}$ is the standard Hamiltonian for electrons
in a disordered conductor and $\hat H_{\rm int}$ describes
interaction between the particle and the electronic environment.
 Fluctuating electrons in
this environment produce stochastic electromagnetic field $V$
described by the equilibrium correlator
\begin{equation}
\langle VV \rangle =
T\sum_{\omega_n}\int\frac{d^3k}{(2\pi)^3}\frac{4\pi}{k^2\epsilon
(i|\omega_n|,k)}e^{-i\omega (\tau -\tau')+ i\bbox{kX}}, \label{VV}
\end{equation}
where $\omega_n=2\pi nT$ is the Matsubara frequency, $\epsilon
(\omega , k)$ is the dielectric susceptibility of the environment
and $\bbox{X}=\bbox{R}(\tau )-\bbox{R}(\tau')$. Similarly to Refs.
\onlinecite{Paco,GHZ} we will model the environment by 3d
diffusive electron gas with
\begin{equation}
\frac{1}{\epsilon (\omega , k)}\approx \frac{-i\omega +Dk^2}{4\pi \sigma},
\end{equation}
where $\sigma$ is the Drude conductivity of this gas, $D=v_Fl/3$
is the electron diffusion coefficient and $l$ is the electron
elastic mean free path. Interaction between the particle on a ring
and fluctuating electrons in the environment is described by the
standard Coulomb term
\begin{equation}
\hat H_{\rm int}=e\hat V. \label{eV}
\end{equation}

In what follows we will assume that the whole system remains in
thermodynamic equilibrium at a temperature $T$. Our first and
standard step is to integrate out all environmental degrees of
freedom effectively described by the collective variable $V$. In
the limit of weakly disordered environment $k_Fl \gg 1$ to be
analyzed below fluctuations of the field $V$ can be considered
Gaussian. In this case integration over this field is carried out
exactly \cite{Paco,GHZ}. After that one arrives at the grand
partition function of the system expressed as a single path
integral over the angle variable $\theta (\tau )$:
\begin{eqnarray} {\cal Z}\equiv\sum_{m=-\infty}^{\infty}e^{2\pi i m\phi_x}{\cal Z}_m =
\qquad\qquad \qquad\qquad\qquad\nonumber\\=\sum_{m=-\infty}^{\infty}\int_0^{2\pi}d\theta_0\int_0^{2\pi m}{\cal
D}\theta \exp (2\pi im\phi_x-S[\theta ]). \label{Z}
\end{eqnarray}
Here we defined $\beta=1/T$, $\phi_x=\Phi_x/\Phi_0$ and the
winding-number-projected partition functions ${\cal Z}_m$. The
first term in the exponent in Eq. (\ref{Z}) takes care of the
magnetic flux while the second term $S[\theta ]$ describes the
effective action for our interacting particle on a ring. This
action consists of two terms,
\begin{equation}
S[\theta ]=S_0[\theta ]+S_{\rm int}[\theta ]. \label{act}
\end{equation}
The term
\begin{equation}
S_0[\theta ]=\int_0^{\beta}d\tau
\left[\frac{1}{4E_C}\left(\frac{\partial \theta}{\partial
    \tau}\right)^2+U_0(1-\cos (\kappa \theta ))\right]
\label{part}
\end{equation}
defines the action for a particle in the absence of the
environment. This action is identical to one for the Josephson
junction \cite{SZ} where $E_C$ plays the role of the charging
energy. The term $S_{\rm int}$ describes the effect of interaction
between the particle and the environment. For our model it has the
form \cite{Paco,GHZ}
\begin{eqnarray}
S_{\rm int}[\theta ]=\alpha\int_0^{\beta}d\tau
\int_0^{\beta}d\tau'\frac{\pi^2T^2K(\theta (\tau )-\theta (\tau'))}
{\sin^2 [\pi T(\tau -\tau')]},
\label{actionD}\\
K(z)=1-\frac{1}{\sqrt{4r^2\sin^2(z/2)+1}},
\label{F}
\end{eqnarray}
where the constant $\alpha =3/(8k_F^2l^2)$ effectively controls
the interaction strength in our model and $r=R/l$. Note that the
integral in Eq. (\ref{actionD}) is understood as a principal
value. The formal divergence at $\tau=\tau'$ is regularized by
requiring $K(0)=0$ which explains the origin of the first term in
(\ref{F}). For the sake of physical consistency of our model below
we will set $1/k_F \ll l \ll 2\pi R/\kappa$, where the first
inequality just means that interaction should remain weak, $\alpha
\ll 1$, while the second one implies that the distance between the
neighboring potential minima should be much larger than $l$.
Accordingly, the parameter $r=R/l$ obeys the inequality
\begin{equation}
r\gg \kappa /2\pi. \label{lr}
\end{equation}
Note that it can also be convenient to rewrite the function $K(z)$
in terms of the Fourier series
\begin{equation}
K(z)=\sum_{n}a_n\sin^2\left[\frac{nz}2\right],
\label{Fourier}
\end{equation}
where the Fourier coefficients are
$a_n \sim (2/\pi r)\ln (r/n)$ for $1\leq n \lesssim r$ and
$a_n \approx 0$ otherwise.

\section{Persistent current in the absence of dissipation}

Let us first evaluate persistent current in our system in the
absence of interactions, i.e. we set the interaction constant
equal to zero $\alpha =0$ everywhere in this section. In what
follows we will mainly be interested in sufficiently large values
of the ring radius $R$. Hence, without loss of generality one can
consider the limit $E_C \ll U_0$. Below we will perform our
calculation for a somewhat more stringent condition
\begin{equation}
\kappa^2E_C \ll U_0 \label{RRR}
\end{equation}
which simplifies our analysis but is by no means important for any
of our key conclusions.

In the tight binding limit (\ref{RRR}) and at sufficiently low
temperatures the particle is located at the bottom of one of the
potential wells (see Fig. \ref{f1}), i.e. in the vicinity of the points
$\theta =2\pi p/\kappa$, where $0< p \leq \kappa$ is an integer
number. Accordingly, in Eq. (\ref{Z}) one should substitute
\begin{equation}
\int_0^{2\pi}d\theta_0 \to \sum_{p=1}^{\kappa}\int d\theta_0\delta
(\theta_0-2\pi p/\kappa ).
\end{equation}
The particle can move around the ring only by hopping between the
neighboring minima $\theta =2\pi p/\kappa$ and $\theta =2\pi (p\pm
1)/\kappa$ of the periodic potential $U(\theta )$. Each of these
tunneling events is described by the well known instanton (kink)
trajectory
\begin{equation}
\tilde \theta (\tau )= \frac{4}{\kappa}\arctan (\exp (\omega \tau
)) \label{inst}
\end{equation}
and corresponds to the tunneling rate $\Delta /2$, where
\begin{equation}
\Delta =8\left(\frac{\kappa U_0E_C}{\pi}\right)^{1/2}
\left(\frac{2U_0}{E_C}\right)^{1/4}\exp
\left(-\frac{4}{\kappa}\sqrt{\frac{2U_0}{E_C}}\right) \label{rate}
\end{equation} and $\omega =\kappa \sqrt{2U_0E_C}$. In
order to evaluate the grand partition function
\begin{eqnarray}
{\mathcal Z} \sim \sum_{p=1}^{\kappa}  \langle p|e^{-\beta \hat
H_0}|p\rangle \label{gpf}
\end{eqnarray}
it is necessary to sum over all possible tunneling events of the
particle between all potential minima to all orders in $\Delta$.
The minimum number of such hops should be equal to $m\kappa$ for
any trajectory corresponding to the winding number $m$. Taking
into account that effective duration of each tunneling event is
$\sim \omega^{-1}$ and that the total imaginary time span equals
to $\beta \equiv 1/T$ we can distinguish two different limits. In
the limit $m\kappa\omega^{-1}\ll\beta$ the average distance
between instantons is large as compared to their typical size,
i.e. in this case we are dealing with dilute instanton gas. In the
opposite limit $m\kappa\omega^{-1}\gg\beta$ instantons are very
close to each other and essentially merge forming a single
trajectory. Below we will also demonstrate that in the above
conditions it suffices to set $m =1$.

\subsection{Dilute instanton gas}

We begin with the low temperature limit $T \ll \omega/\kappa$. To
proceed let us evaluate a somewhat more general than in Eq.
(\ref{gpf}) matrix element $\langle p_1|e^{-\beta \hat
H_0}|p_2\rangle$. For this purpose we consider multi-instanton
trajectories
\begin{equation}
\Theta (\tau )= \frac{2\pi p_1}{\kappa}+\sum_j\nu_j\tilde \theta
(\tau-\tau_j), \label{multiinst}
\end{equation}
where $\nu_j=\pm 1$ and $\tau_j$ are respectively the topological
charges and collective coordinates of instantons and $\tilde\theta
(\tau)$ is defined in Eq. (\ref{inst}). The trajectory
(\ref{multiinst}) describes tunneling of the particle between the
states $|p_1\rangle$ and $|p_2\rangle$ after $m$ winds around the
ring provided we fix
\begin{equation}
\sum_j\nu_j=n_2-n_1=p_2-p_1+m\kappa , \label{rest}
\end{equation}
i.e. we consider configurations containing totally $n_1+n_2$
instantons corresponding to $n_1$ hops clockwise and $n_2$ hops
counterclockwise.  Taking into account all possible tunneling
events restricted by the condition (\ref{rest}) and summing over
all winding numbers $m$, we obtain
\begin{eqnarray}
\label{allor}
  \langle p_1|e^{-\beta \hat H_0}|p_2\rangle &=& \left(\frac{\omega}{\pi}\right)^{1/2}
  \sum_{m=-\infty}^{\infty} e^{2\pi
im\phi_x-\frac{\beta \omega}{2}}\\ &&
\times\sum_{n_1,n_2=0}^{\infty}\frac{(\beta
\Delta/2)^{n_1+n_2}}{n_1!n_2!}\delta_{n_1-n_2,p_1-p_2-m\kappa}.
\nonumber
\end{eqnarray}
Making use of the integral representation for the Kronecker symbol
\begin{equation}
 \delta_{j,k}=\int\limits_{0}^{2\pi}\frac{dy}{2\pi}e^{iy(j-k)},
\end{equation}
after performing a summation over $m$ with the aid of Poisson's resummation
formula
\begin{equation}
 \sum_{m=-\infty}^{\infty} e^{2\pi i m x}=\sum_{k=-\infty}^{\infty}\delta(x-k)
\end{equation}
we obtain
\begin{eqnarray}
 \label{sum}
 \langle p_1|e^{-\beta \hat
  H_0}|p_2\rangle &=& \left(\frac{\omega}{\pi\kappa^2}\right)^{1/2}e^{-\frac{\beta
    \omega}{2}} \\ &&
\times\sum_{j=1}^{\kappa}
e^{i\frac{2\pi(j+\phi_x)(p_1-p_2)}{\kappa}+\beta \Delta
\cos\left(\frac{2\pi(j+\phi_x)}{\kappa}\right)}. \nonumber
\end{eqnarray}
This formula allows to easily recover the low-lying energy levels
of our problem which contain all necessary information in order to
evaluate PC. For instance, the ground state energy of the particle
$E_0(\phi_x)$ is obtained in a standard way by taking the limit
$T=1/\beta \to 0$ in Eq. (\ref{sum}) which yields
\begin{equation}
E_0(\phi_x
)=\frac{\omega}{2}-\Delta\cos\left(\frac{2\pi\phi_x}{\kappa}\right)\label{gse}
\end{equation}
for $-1/2 <\phi_x <1/2$. Eq. (\ref{gse}) should be continued
periodically outside this interval. This expression determines the
periodic dependence (with period equal to the flux quantum
$\Phi_0$) of the ground state energy on the magnetic flux
$\Phi_x$. In the limit $\kappa \gg 1$ this dependence reduces to a
set of parabolas
\begin{equation}
E_0(\phi_x)-\frac{\omega}{2}\simeq
\frac{2\pi^2\Delta}{\kappa^2}{\rm min}_n(\phi_x -n)^2.
\label{gse1}
\end{equation}

\begin{figure}[t]
\includegraphics[width=3.375in]{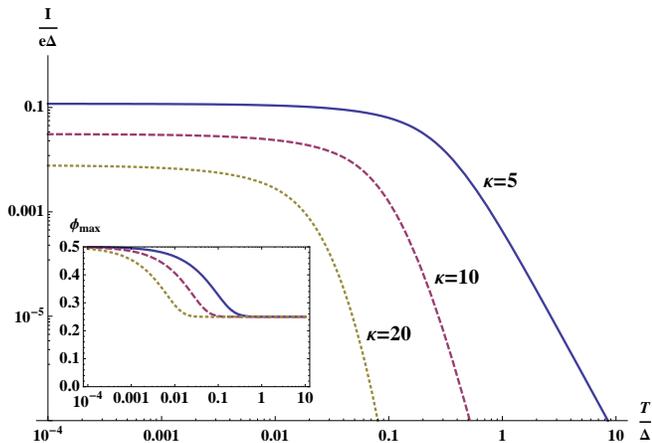}
\caption{\label{f2} (Color online) Maximum value of PC as a
function of temperature for different values of $\kappa$. Inset:
The magnetic flux value $\phi_{max}$ at which PC reaches its
maximum value as a function of $T/\Delta$.}
\end{figure}

Turning back to the grand partition function (\ref{gpf}), from Eq.
(\ref{sum}) we find
\begin{equation}
{\mathcal Z} \sim \sum_{j=1}^{\kappa} e^{\beta
\Delta\cos\left(\frac{2\pi(j+\phi_x)}{\kappa}\right)}.
\end{equation}
PC can now be easily obtained from the general formula
\begin{equation}
I =-\frac{eT}{2\pi}\frac{\partial}{\partial\phi_x}\ln{\mathcal Z},
\label{tok}
\end{equation}
which yields diamagnetic current
\begin{equation}
I =\frac{e\Delta}{\kappa}
\frac{\sum_{j=1}^{\kappa}\sin\left(\frac{2\pi(j+\phi_x)}{\kappa}\right)
e^{\beta
\Delta\cos\left(\frac{2\pi(j+\phi_x)}{\kappa}\right)}}{\sum_{j=1}^{\kappa}
e^{\beta \Delta\cos\left(\frac{2\pi(j+\phi_x)}{\kappa}\right)}}.
\label{genexp}
\end{equation}
This expression fully determines PC in the ring at temperatures $T
\ll \omega/\kappa$ and in the absence of interactions with
dissipative environment. The dependence of the maximum PC value on
temperature is depicted in Fig. \ref{f2}.

Eq. (\ref{genexp}) is further simplified at temperatures above and
below the interlevel distance $\sim \Delta/\kappa^2 $. In the
limit $T \gg \Delta /\kappa^2$ the leading contribution to the
partition function is defined by configurations with minimal
number of instantons. Hence, in the sum over winding numbers in
Eq. (\ref{Z}) it is sufficient to keep only the terms with
$m=0,\pm1$ terms. For the term with $m=0$ it is necessary to sum
over all configurations, whereas for the case $m=\pm1$ only
configurations with $\kappa$ instantons contribute. After some
algebra we get
\begin{equation}
  {\cal Z}_0=\kappa e^{-\beta\omega/2}{\rm I_0}(\beta\Delta),
\end{equation}
where ${\rm I_0}(x)$ is the modified Bessel function of imaginary
argument. For $m=\pm1$ we find
\begin{equation}
    {\cal Z}_{\pm1}=\kappa
    \frac{(\beta\Delta)^\kappa}{2^\kappa\kappa!}e^{-\beta\omega/2}.
\end{equation}
 As a result we obtain
 \begin{equation}
I=I_{C0}(T)\sin(2\pi\phi_x), \label{sin}
\end{equation}
where
\begin{equation}
I_{C0}(T)=\frac{e
\Delta^\kappa}{(2T)^{\kappa-1}\kappa!{\rm
 I_0}\left(\Delta/T\right)}.
 \label{highT}
\end{equation}

At low temperatures $T \ll \Delta /\kappa^2$ Eq. (\ref{genexp})
reduces to a simple formula
\begin{equation}
I=\frac{e\Delta}{\kappa}\sin\left(\frac{2\pi\phi_x}{\kappa}\right),
\quad -1/2 < \phi_x \leq 1/2, \label{lowT}
\end{equation}
which also trivially follows from Eq. (\ref{gse}). This formula
demonstrates that at $T=0$ the magnitude of PC in our system is
proportional to $\Delta$ while its flux dependence deviates from
the simple sinusoidal form for all $\kappa >1$. In particular, in
the limit $\kappa \gg 1$ this dependence approaches a sawtooth one
\begin{equation}
I=I_{C0}{\rm min}_n(\phi_x -n), \quad I_{C0}=2\pi e\Delta /
\kappa^2. \label{lk}
\end{equation}

Comparing the expressions for PC derived above with those for a
free particle on a ring \cite{Paco,GHZ,HlD,CH,KH} we observe that
the main physical difference between these two models is the
presence of two distinct energy scales -- $\omega$ and $\Delta$ --
in our case whereas only one energy scale $E_C$ remains in the
limit $U_0=0$. Otherwise significant features of the effect are
essentially the same in both models. Indeed, at high temperatures
($T> \Delta $ here and $T
> E_C$ for $U_0=0$) the dependence of PC on the magnetic flux
$\Phi_x$ is sinusoidal with the period $\Phi_0$ and its amplitude
decreases with increasing $T$, cf. our Eq. (\ref{highT}) with Eq.
(11) in Ref. \onlinecite{GHZ}. In the low temperature limit ($T
\ll \Delta /\kappa^2$ here and $T \ll E_C$ for $U_0=0$) the
dependence $I(\phi_x)$ strongly deviates from sinusoidal (except
for a special case $\kappa =1$). In the limit $\kappa \gg 1$ the
sawtooth dependence (\ref{lk}) is identical to that for the case
$U_0=0$ where one has $I_{C0} \sim E_C$.

Finally, let us compare the dependence of the PC amplitude
$I_{C0}$ on the ring radius $R$ obtained in these two cases.
Provided  the parameter $\kappa$ is fixed and does not change with
$R$,  Eqs. (\ref{lk}) and (\ref{rate}) yield exponential decay of
$I_{C0}$ with increasing $R$ since $E_C \propto 1/R^2$. This
exponential decay is due to the fact that for larger $R$ the
potential profile changes in a way that the particle should tunnel
at a longer distance $2\pi R/\kappa$ between the two neighboring
states $|p\rangle$ and $|p\pm 1\rangle$. Alternatively, one can
keep the distance between the adjacent potential minima $\theta
=2\pi p/\kappa$ and $\theta =2\pi (p\pm 1)/\kappa$ unchanged while
increasing $R$. This is achieved by varying the parameter $\kappa$
with $R$ as $ \kappa \propto R$. Under this condition the
tunneling rate $\Delta$ (\ref{rate}) becomes independent of $R$
and the magnitude of PC $I_{C0}$ (\ref{lk}) decreases with
increasing $R$ as $I_{C0} \propto 1/R^2$ exactly as in the case
$U_0=0$. Note, however, that due to the restriction (\ref{RRR})
this regime can apply only at not too large values of $R$.

\subsection{Merged instantons}

Now let us turn to the case of higher temperatures $\omega/\kappa
\ll T \ll \omega$ which can be realized in the limit $\kappa \gg
1$. In this case the leading contribution to the partition
function originates from one multi-instanton trajectory. This
trajectory of merged instantons $\Theta^{(m)}(\tau)$ can easily be
evaluated due to the presence of the integral of motion which is
the classical energy ${\cal E}_m$ corresponding to the winding
number $m$.  This energy is fixed by the periodic boundary
condition
 \begin{equation}
   \frac{1}{T}=\sqrt{\frac{MR^2}{2}}\int\limits_0^{2\pi |m|}\frac{d\theta}{\sqrt{{\cal E}_m+U_0(1-\cos(\kappa\theta))}}
 \end{equation}
and the trajectory $\Theta^{(m)}(\tau)$ is obtained from the
equation
 \begin{equation}
     \tau-\tau_0=\sqrt{\frac{MR^2}{2}}\int\limits_0^{\Theta^{(m)}}\frac{d\theta}{\sqrt{{\cal
     E}_m+U_0(1-\cos(\kappa\theta))}}.
 \end{equation}
Making use of the standard quasiclassical technique one arrives at
the following expression for the partition function:
\begin{equation}
{\cal Z}=\sum\limits_{m=-\infty}^\infty\left(-\frac{\partial^2
S({\cal E}_m)}{\partial {\cal E}_m^2}\right)^{-1/2} \frac{e^{2\pi
i m\phi_x-|m|S({\cal E}_m)+{\cal E}_m/T}}{T\sqrt{2\pi|m|^3}},
\end{equation}
where
\begin{equation}
S({\cal E})=\frac{1}{\sqrt E_C}\int\limits_0^{2\pi}\sqrt{{\cal E}+U_0(1-\cos(\kappa\theta))}d\theta
\end{equation}
is the classical action. As before, the leading contribution to
this partition functions comes from the lowest winding numbers
$m=0,\pm1$. For $m=0$ one recovers the oscillator-like expression
\begin{equation}
{\cal Z}_0=\frac{\kappa}{2\sinh\frac{\omega}{2T}}\approx\kappa
e^{-\frac{\omega}{2T}},
\end{equation}
whereas for $m=\pm1$ the instanton trajectory approaches the
straight line in which case the integrals can be easily evaluated
and yield
\begin{equation}
\mathcal Z_{\pm 1}=\kappa\sqrt{\frac{2\pi U_0
T}{\omega^2}}e^{-\frac{2\pi^2U_0 T\kappa^2}{\omega^2}}.
\end{equation}
With the aid of these results one again arrives at the expression
for PC in the form (\ref{sin}) with
\begin{equation}
I_{C0}(T)=2eT\sqrt{\frac{2\pi U_0
T}{\omega^2}}\exp\left(\frac{\omega}{2T}-\frac{2\pi^2U_0
T\kappa^2}{\omega^2}\right).
\label{vhighT}
\end{equation}

\begin{figure}[t]
\includegraphics[width=3.375in]{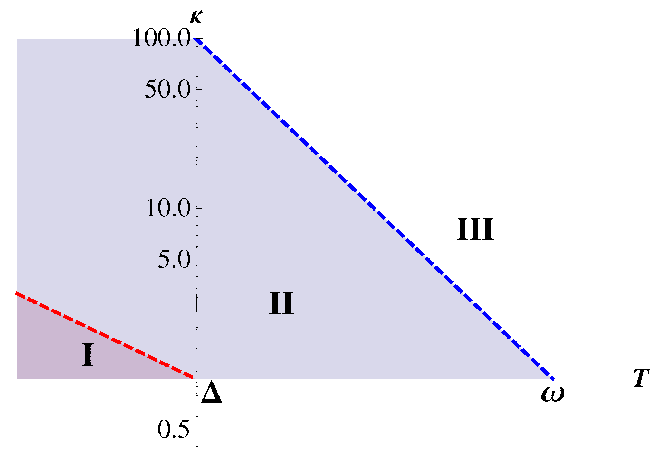}
\caption{\label{f3} (Color online) Different regimes for PC in the
non-interacting case. The regions I (lowest $T$) and II (higher
$T$) are described within the dilute instanton gas approximation
(Eqs. (\ref{lowT}) and (\ref{sin}), (\ref{highT}) respectively),
whereas the high temperature region III corresponds to the case of
merged instantons, Eqs. (\ref{sin}), (\ref{vhighT}). The dashed
lines $T\approx \Delta/\kappa^2$ and $T \approx \omega/\kappa$
indicate the crossover between these regimes. }
\end{figure}

Different regimes considered in this section can also be
summarized graphically by means of the diagram depicted in Fig.
\ref{f3}.

\section{Effect of electron-electron interactions}

\subsection{Renormalization of the tunneling amplitude}

Now let us turn on interactions and analyze the effect of
fluctuations in a dissipative environment. To this end we
again employ the above instanton technique. Evaluating the path
integral in (\ref{Z}) in the limit (\ref{RRR}) and for
$T\ll\omega/\kappa$ we follow the same scheme and substitute the
trajectory (\ref{multiinst}) describing quantum tunneling of the
particle between different potential minima into the full
effective action (\ref{act}). As before, let us fix the winding
number equal to $m$. Then we again arrive at configurations of
totally $k$ instantons restricted by the condition (\ref{rest})
where we now also set $p_1=p_2=p$. Evaluating the interaction term
$S_{\rm int}$ (\ref{actionD}), (\ref{F}) on multi-instanton
trajectories (\ref{multiinst}) after some algebra (see Appendix A)
we obtain
\begin{eqnarray}
S_{\rm int}[\Theta (\tau )]=4\pi \alpha rk/\kappa\qquad\qquad\qquad\nonumber \\
 -2\alpha\sum\limits_{a,b=1,\ a<b}^{k}\nu_a \nu_b
g(\varphi_{ab})
   \ln\left[\frac{\sin[\pi T(\tau_b-\tau_a)]}{\pi
   T\omega^{-1}}\right],
\label{instint}
\end{eqnarray}
where
\begin{equation}
   g(\varphi)=K(\varphi+2\pi/\kappa)+K(\varphi-2\pi/\kappa)-2K(\varphi)
\end{equation}
and
\begin{equation}
\varphi_{ab}=\frac{2\pi}{\kappa}\sum\limits_{j=a}^{b}\nu_j-\frac{\pi}{\kappa}(\nu_b+\nu_a).
\end{equation}
We observe that the interaction term (\ref{instint}) consists of
two different contributions. One of them, $4\pi \alpha rk/\kappa$,
describes interaction-induced suppression of quantum tunneling of
the angle variable $\theta$ between different potential minima.
This term yields effective $r$-dependent renormalization of the tunneling
amplitude
\begin{equation}
\Delta\rightarrow\Delta_r=\Delta e^{-4\pi\alpha r/\kappa}.
\label{dren}
\end{equation}
The remaining contribution in Eq. (\ref{instint}) describes
logarithmic interaction between different instantons which occurs
for $\kappa \geq 2$ due to the presence of a dissipative
environment. This logarithmic interaction is absent for $\kappa
=1$ in which case $g(\varphi ) \equiv 0$.

At not too low temperatures $T\gg \Delta_r/\kappa^2$
inter-instanton interactions just provide further renormalization
of the tunneling amplitude $\Delta_r\rightarrow \Delta_r(1+2\alpha
K(2\pi/\kappa)\ln(2\pi T/\omega))$. One can also write down the
renormalization group (RG) equation
\begin{equation}
\frac{d\Delta_r}{d\ln\omega}=2\alpha K(2\pi/\kappa)\Delta_r,
\label{rg}
\end{equation} which yields both $r-$ and $T-$dependent
renormalized tunneling amplitude of the form
\begin{equation}
\Delta_R=\Delta_r\left(\frac{ T}{\omega}\right)^{2\alpha
K(2\pi/\kappa)}. \label{renht}
\end{equation}
At even lower temperatures $T\ll\Delta_r/\kappa^2$ interactions
again yield renormalization of the tunneling amplitude which now
becomes $\Delta_r\rightarrow \Delta_r\left(1+2\alpha
K(2\pi/\kappa)\ln\frac{2\Delta_r e^\gamma}{\omega}\right)$. The
corresponding RG equation (\ref{rg}) remains the same but relevant
energy scale now becomes
\begin{equation}
\Delta_R=\Delta_r\left(\frac{\Delta_r}{\omega}\right)^{\frac{2\alpha
K(2\pi/\kappa)}{1-2\alpha K(2\pi/\kappa)}}. \label{renlt}
\end{equation}
Eq. (\ref{renlt}) -- together with Eq. (\ref{dren}) -- defines the
renormalized tunneling amplitude in the limit $T \to 0$ and for
$\kappa \geq 2$. In the particular case $\kappa=2$  our results
reduce to those established for the so-called spin-boson model
with Ohmic dissipation \cite{Leggett,Weiss}.

We also note that Eq. (\ref{renlt}) is formally applicable for
$\alpha < 1/(2K(2\pi/\kappa ))$, while for larger values of the
interaction strength $\alpha$ and at $T=0$ the tunneling amplitude
is renormalized to zero, $\Delta_R = 0$. This is the consequence
of the quantum dissipative phase transition which -- similarly to
other models \cite{SZ,Leggett,Weiss} --  occurs at the critical
interaction strength $\alpha_c =1/(2K(2\pi/\kappa ))\simeq 1/2$ and
implies localization of a quantum particle in one of potential
wells at any $\alpha$ exceeding the critical value $\alpha_c$.
Accordingly, no PC can flow at $T=0$ and $\alpha \geq
1/(2K(2\pi/\kappa ))$. This formal conclusion, however, does not
appear to be of substantial physical significance, since the
applicability range of our model is restricted to small values of
$\alpha \ll 1$.

Finally, we should point out that, employing a regular
perturbation theory in $\alpha$, already in the first order one
recovers additional terms which cannot be captured within the RG
equation (\ref{rg}). The corresponding analysis is presented in
Appendix B.

\subsection{Suppression of PC by interactions}

It turns out that the behavior of PC may be quite different
depending both on temperature and on the parameter $\kappa$.
Therefore, it is appropriate to distinguish several different
cases.

\subsubsection{One potential minimum $\kappa =1$}
As we already discussed, in the case $\kappa =1$ logarithmic interaction
between instantons is absent, and the only effect of interactions is
$r-$dependent renormalization of the tunneling amplitude
(\ref{dren}). Accordingly, for PC in this case we obtain
  \begin{equation}
I=e\Delta e^{-4\pi\alpha r}\sin (2\pi\phi_x).
\label{k1int}
\end{equation}
This result is valid at all temperatures in the range $T \ll
\omega$. It demonstrates that for $\kappa =1$ Coulomb interaction
yields exponential suppression of PC down to $T=0$ provided the
ring perimeter $2\pi R$ exceeds an effective dephasing length
\begin{equation}
L_{\varphi} \sim l/\alpha, \label{dephl}
\end{equation}
which is set by the effective interaction strength $\alpha$ and does not
depend on temperature. Note that exactly the same length scale was found
in the absence of the periodic potential $U_0=0$ in Ref. \onlinecite{GHZ}.

\subsubsection{Not too low temperatures and $\kappa \geq 2$}

In the case $\kappa \geq 2$ instantons interact logarithmically
and -- in addition to (\ref{dren}) -- at not too low temperatures
the tunneling amplitude is renormalized according to Eq.
(\ref{renht}). Combining these two equations and substituting the
renormalized tunneling amplitude $\Delta_R$ into Eq. (\ref{highT})
instead of $\Delta$ we obtain
\begin{equation}
 I=I_{C}(T)\sin(2\pi\phi_x),
\label{sin11}
\end{equation}
where
\begin{equation}
I_{C}(T)= \left\{\begin{array}{lc}I_{C0}(T) \left(\frac{ T}{\omega}\right)^{2\alpha \kappa
K(2\pi/\kappa)}e^{-4\pi\alpha r}, & T\ll \omega/\kappa ,\\
I_{C0}(T)e^{-4\pi\alpha r}, & T\gg \omega/\kappa
 \end{array}\right.
\label{highTIc}
\end{equation}
and $I_{C0}(T)$ in the corresponding limits is defined
respectively in Eq. (\ref{highT}) and (\ref{vhighT}). In this case
we again recover the same exponential suppression of PC at ring
perimeters $2\pi R \gtrsim L_{\varphi}$ with {\it temperature
independent} dephasing length defined in Eq. (\ref{dephl}). At the
same time, the pre-exponent in the expression for $I_C$ depends on
temperature as a power law $I_C(T) \propto T^{-\mu}$ with $\mu
=\kappa (1-2\alpha K(2\pi/\kappa ))-1$. For $\kappa \geq 2$ and
small values of the interaction strength $\alpha \ll 1$ we have
$\mu >0$, i.e. $I_C(T)$ grows with decreasing temperature. This
growth, though somewhat weaker than in the non-interacting case
(\ref{highT}) (since $\mu < \kappa -1$), implies that Eqs.
(\ref{sin11}), (\ref{highTIc}) can be trusted only at $T \gtrsim
\Delta_R/\kappa^2$ whereas at even lower temperatures one expects
a crossover to a different regime to be discussed below. We also
note that qualitatively similar behavior of PC at not too low
temperatures follows from the numerical analysis \cite{GHZ} of the
model with $U_0=0$.

\subsubsection{Zero temperature limit and  $\kappa \geq 2$}

In order to evaluate PC in the limit $T \ll \Delta_r/\kappa^2$ we
will make use of  Eq. (\ref{B14}) for the free energy. Combining
this expression with Eq. (\ref{tok}) and observing that the
difference $\Delta_R-\Delta_r \sim \alpha \ll 1$ in the last term
in Eq. (\ref{B14}) can be safely neglected within the accuracy of
our calculation (since it only produces extra terms $\sim
\alpha^2$), we obtain
\begin{widetext}
\begin{equation}
I=\frac{e\Delta_R}{\kappa}\left[\sin\left(\frac{2\pi\phi_x}{\kappa}\right)
-\alpha\sum\limits_{n=-r}^ra_n\sin\left(\frac{\pi
n}{\kappa}\right)\cos\left(\frac{2\pi\phi_x+\pi
n}{\kappa}\right)\ln\left|\sin\left(\frac{\pi
n}{\kappa}\right)\sin\left(\frac{2\pi\phi_x+\pi
n}{\kappa}\right)\right|\right], \label{PCT0}
\end{equation}
\end{widetext}
where $\Delta_R$ is defined in Eqs. (\ref{renlt}), (\ref{dren}).
This result allows to make the following observations.

Firstly, taking into account the dependence of $\Delta_R$ on $r$
we conclude that exactly at $T=0$ and at sufficiently large ring
perimeters PC is exponentially suppressed as
\begin{equation}
I \propto \exp \left(-\frac{4\pi \alpha r}{\kappa (1-2\alpha K(2\pi/\kappa
  ))}\right),
\label{dephatzero}
\end{equation}
i.e. in this case for any fixed value $\kappa$ we can define an effective
zero temperature dephasing length
\begin{equation}
\tilde L_\varphi =L_\varphi\kappa (1-2\alpha K(2\pi/\kappa )).
\label{tildeph}
\end{equation}
This result demonstrates that in the limit of weak interactions $\alpha \ll 1$
the effective length (\ref{tildeph}) turns out to be approximately
$\kappa$ times longer than $L_{\varphi}$, i.e.
$\tilde L_\varphi \approx l/\alpha \kappa$.  Should, however, we assume that
$\kappa \propto r$, no finite dephasing length could be defined from Eq.
(\ref{dephatzero}), although also in this case even for small $\alpha$
PC can suffer exponentially strong suppression by
electron-electron interactions due to the condition (\ref{lr}).

Secondly, the result (\ref{PCT0}) demonstrates that at $T \to 0$
the effect of electron-electron interactions on PC does not just
reduce to renormalization of the tunneling amplitude $\Delta \to
\Delta_R$. We observe that Eq. (\ref{PCT0}) also contains
additional terms evaluated here within the first order
perturbation theory in $\alpha$. This $r$-dependent first order
contribution turns out to be singular at values of $\phi_x$ close
to $\pm 1/2$ and at all other half-integer numbers which indicates
insufficiency of the first order perturbation theory, at least for
such values of $\phi_x$.  Higher order terms of the perturbation
theory in the interaction may contain similar (or even stronger)
singularities and, on top of that, may grow with increasing $r$.
Hence, at $T \to 0$ no perturbation theory in $\alpha$ can in
general be trusted, in particular at sufficiently large $r$. A
more detailed analysis of this issue is beyond the scope of the
present paper. Here we only conjecture that such analysis might
yield an additional dependence of PC on the ring radius $r$ not
accounted for in the expression for $\Delta_R$. It is quite likely
that such kind of $r$-dependence of PC at $T=0$ was also observed
within a numerical treatment \cite{GHZ} in the case $U_0=0$.
Additional support for this conjecture is provided by the exact
solution presented below.

\subsubsection{Toulouse limit}

As we already pointed out, for $\kappa=2$ our problem is exactly
mapped onto the well known spin-boson model with Ohmic dissipation
\cite{Leggett,Weiss}. In this case interaction between instantons
is given by Eq. (\ref{intk2}) and the grand partition function
reads (see also Appendices A and B)
\begin{widetext}
\begin{equation}
Z[z;\beta]=\sum\limits_{n=0}^\infty\left(\Delta_r\cos
z\right)^{2n}\int\limits_0^\beta d\tau_1\int\limits_{\tau_1}^\beta
d\tau_2...\int\limits_{\tau_{2n-1}}^{\beta}d\tau_{2n}\exp\left(4\alpha\sum\limits_{j<k=1}^{2n}(-1)^{j+k}\ln\left(\frac{\sin(\pi
T(\tau_{k}-\tau_{j})}{\pi T\omega^{-1}}\right)\right).
\label{sb}
\end{equation}
\end{widetext}
In order to non-perturbatively evaluate PC at all values of $r$
and at all temperatures including $T=0$ we can profit from the
exact solution known for the particular value of the interaction
strength $\alpha=1/4$, the so-called Toulouse limit.

Introducing the parameter
$\frac{1}{2\pi\xi(z)}=\frac{\Delta_r^2\cos^2z}{T\omega}$ one can
conveniently write down the exact expression for the partition
function (\ref{sb}) with $\alpha=1/4$ in the form
\begin{equation}
  Z[z;\beta]=\exp\left[\int\limits_{\pi T/\omega}^\infty dx\frac{1-e^{-\frac{x}{2\pi\xi(z)}}}{x\sinh
  x}\right].
\end{equation}
This result allows to immediately establish PC which reads
\begin{equation}
  I=\frac{e\Delta_r^2}{2\omega}\sin(2\pi\phi_x)\int\limits_{\frac{\pi T}{\omega}}^\infty dx
  \frac{e^{-\frac{x\Delta_r^2}{\omega T}\cos^2(\pi\phi_x)}}{\sinh
  x}.
\end{equation}
At not too low temperatures $\omega\gg T\gg\Delta_r$ we obtain
\begin{equation}
  I=\frac{e\Delta^2}{2\omega}e^{-\pi r}\ln\frac{2\omega}{\pi
  T}\sin(2\pi\phi_x).
\end{equation}
In low temperature limit $T\rightarrow 0$ one finds
\begin{equation}
  I=\frac{e\Delta_r^2}{2\omega}\sin(2\pi\phi_x)\int\limits_{\pi}^\infty dy
  \frac{e^{-\frac{y\Delta_r^2}{\omega^2}\cos^2(\pi\phi_x)}}{y}.
\label{52}
\end{equation}
Evaluating the integral in (\ref{52}) we
arrive at the final result
\begin{equation}
  I=\frac{e\Delta^2 }{2\omega}e^{-\pi
  r}\left[\pi r+\ln\left(\frac{\omega^2e^{-\gamma}}{\pi\Delta^2\cos^2(\pi\phi_x)}\right)\right]\sin(2\pi\phi_x).
\label{tlt0}
\end{equation} We observe that both at non-zero
temperatures and exactly at $T=0$ the above exact expressions for
PC demonstrate its exponential suppression at ring perimeters
exceeding $L_{\varphi}=\tilde L_{\varphi}$ (these two length
scales coincide for $\kappa =2$ and $\alpha =1/4$, see Eq.
(\ref{tildeph})). In addition, the result (\ref{tlt0})
demonstrates that at $T=0$ (a) the dependence of PC on $r$
deviates from purely exponential (which is in agreement with our
above conjecture) and (b) the dependence of PC on $\phi_x$
deviates from purely sinusoidal and contains the logarithmic
singularity at half-integer values of $\phi_x$, cf. also Eq.
(\ref{PCT0}).

\section{Discussion}

In this paper we proposed a model which allows for detailed
non-perturbative treatment of the effect of electron-electron
interactions on PC in normal nanorings at low temperatures. Our
investigation employs a well-controlled instanton technique and
yields a rich structure of different regimes. The main features
observed within our analysis can be summarized as follows: ($i$)
Coulomb interaction yields $R$-dependent renormalization of the
tunneling amplitude (\ref{dren}) which, in turn, results in
exponential suppression of PC at large enough $R$, ($ii$)
logarithmic interaction between instantons yields additional
renormalization of the tunneling amplitude described by the RG
equation (\ref{rg}) and ($iii$) electron-electron interactions
generate yet additional contributions not captured by Eqs.
(\ref{dren}) and (\ref{rg}), see Eq. (\ref{PCT0}) and Appendix B.
These contributions may become particularly important at $T \to 0$
indicating the failure of the naive perturbation theory in the
interaction at sufficiently large $R$.

Although the effect ($iii$) still requires additional
non-perturbative analysis, already ($i$) and ($ii$) result in
exponential suppression of PC at any temperature including $T \to
0$ for any given $\kappa$ and at ring perimeters exceeding the
dephasing length set by interactions and defined in Eqs.
(\ref{dephl}) and (\ref{tildeph}). Note that the length scale
identical to (\ref{dephl}) also follows from the earlier
non-perturbative analysis \cite{GHZ} developed for the case
$U_0=0$. Thus, similarly to Ref. \onlinecite{GHZ} the decoherence
effect in our model is controlled by the parameter
$$
\alpha r \sim \alpha \sum_{n=1}^rna_n
$$
rather than by $\alpha$ or $\alpha \ln r$ as it was sometimes
suggested in the literature in the case $U_0=0$.

It is worthwhile to stress that exponential dependence of PC on
$R$ of the form $I \propto \exp (-AR)$ by itself does not yet
necessarily imply decoherence. For instance, even in the absence
of interactions  PC $I \propto \Delta$ (\ref{lowT}) can decrease
exponentially with increasing $R$ provided the parameter $\kappa$
is fixed to be independent of the ring radius. Obviously, quantum
coherence is not destroyed in this case. Exponential reduction of
PC with increasing $R$ at $T \to 0$ can also occur in
superconducting nanorings due to proliferation of quantum phase
slips \cite{MLG,AGZ}.  Also in that case the dependence $I \propto
\exp (-AR)$ can be interpreted just as a non-trivial
coordinate-dependent renormalization effect \cite{AGZ}.

An important qualitative difference between nanorings with
dissipation considered in Sec. 4 and the two last examples is that
in our problem dissipation explicitly \textit{violates
time-reversal symmetry} (thus causing genuine decoherence of a
quantum particle), while no such symmetry is violated in
superconducting nanorings \cite{MLG,AGZ} or in the absence of
dissipation (Sec. 3). Hence, quantum coherence remains fully
preserved in the last two cases despite exponential suppression of
PC at large $R$.

One can discriminate between decoherence  and pure renormalization
in a number of ways. For instance, one can drive the system out of
equilibrium and investigate its relaxation by means of a real-time
analysis. This approach was employed in Refs. \onlinecite{GHZ,GSZ}
where a finite dephasing time $\tau_{\varphi}=L_{\varphi}/v$ was
found at $T=0$ with $L_{\varphi}$ defined in Eq. (\ref{dephl}) and
$v$ being the particle velocity. This observation allows to
unambiguously identify quantum decoherence.

Another way \cite{Buttiker} amounts to analyzing fluctuations of
PC in the ground state of an interacting system. For instance, one
can study the correlator $\langle (\hat I -\langle \hat I \rangle
)^2\rangle$, where $\hat I$ is the current operator which
expectation value $\langle \hat I \rangle$ defines PC in the
ground state. Within the model \cite{Buttiker} it was demonstrated
that, while the average PC in the ground state decreases with
increasing interaction strength, its fluctuations {\it increase},
thus implying genuine decoherence rather than pure
renormalization. A similar situation occurs within the model
studied here.

Following \cite{Paco} here we intentionally disregarded Fermi
statistics by suppressing electron exchanges between the ring and
the environment. The question arises if inclusion of the Pauli
principle into the model could alter our main conclusion about
non-vanishing electron decoherence at zero temperature. Golubev
and one of the present authors addressed this issue by developing
two entirely different non-perturbative in the interaction
techniques \cite{GZ1,GZ2}. Both these methods yield the same
conclusion: Although Fermi statistics is crucially important for
other properties of a dirty interacting electron gas, it
practically {\it does not affect} interaction-induced quantum
decoherence at $T \to 0$. Quantitative agreement was demonstrated
\cite{GHZ} between the corresponding results derived for the
models with \cite{GZ1} and without \cite{GHZ} the Pauli principle.

Despite all these developments (including recently obtained {\it
exact} solution of the problem \cite{GZ2}) it is sometimes argued
in the literature that the Pauli principle can preclude from
electron decoherence at $T \to 0$. E.g. this conclusion was
reached on the basis of the first order perturbation theory in the
interaction \cite{AAG}. Insufficiency of such kind of perturbation
theory for the problem in question was repeatedly demonstrated
elsewhere \cite{GZ1,GZ2,GHZ,GZS} and was also observed here within
the model considered. More recently, von Delft and co-authors
\cite{JvD} re-iterated the same incorrect conclusion \cite{AAG}.
Unfortunately the analysis \cite{JvD} is fundamentally flawed
since it violates causality effectively implying that dynamics of
interacting electrons would be affected by photons coming both
from the past and from the future. Hence, the results \cite{JvD}
cannot be considered seriously. For more details on this issue we
refer the reader to the paper \cite{GZ08}.

 \vspace{0.3cm}

\centerline{\bf Acknowledgments}

\vspace{0.3cm}

This work was supported in part by RFBR grant 09-02-00886. A.G.S.
also acknowledges support from the Landau Foundation and from the
Dynasty Foundation.


\begin{widetext}
\appendix
\section{Instanton gas in the presence of interactions}

Let us analyze the effect of electron-electron interactions on the dilute
instanton gas employed in our work. Combining Eqs. (\ref{actionD}) and
(\ref{Fourier}) we can rewrite the dissipative part of the action
describing such interactions within our model in the following form:
\begin{equation}
 S_{\rm int}[\theta]=-\frac{\alpha}{2}\sum\limits_na_n\int\limits_{-\beta/2}^{\beta/2}d\tau
 \int\limits_{-\beta/2}^{\beta/2} d\tau'\frac{\pi^2 T^2 (1-e^{in\theta(\tau)})(1-e^{-in\theta(\tau')})}{\sin^2(\pi
 T(\tau-\tau'))}.
\label{A1}
\end{equation}
Substituting multi-instanton trajectories $\Theta (\tau)$
(\ref{multiinst}) into Eq. (\ref{A1}) and employing the
approximation
\begin{equation}
   1-e^{in\Theta(\tau)}\approx\sum\limits_j e^{\frac{2\pi i n}{\kappa}\sum\limits_{l=1}^{j-1}\nu_l}(1-e^{in\nu_j\tilde\theta(\tau-\tau_j)})\equiv\sum\limits_j C_jf_{nj}(\tau)
\end{equation}
appropriate for well-separated instantons, with the aid of the identity
\begin{equation}
\frac{\pi^2 T^2}{\sin^2(\pi
  T(\tau-\tau'))}=\frac{\partial^2}{\partial\tau\partial\tau'}\ln|\sin(\pi
T(\tau-\tau'))|
\end{equation}
we obtain
\begin{eqnarray}
S_{\rm
int}[\Theta]=-\frac{\alpha}{2}\sum\limits_na_n\sum\limits_{j_1,j_2}C_{j_1}\bar
C_{j_2}
\int\limits_{-\beta/2}^{\beta/2}d\tau\int\limits_{-\beta/2}^{\beta/2}
d\tau' f'_{nj_1}(\tau)\bar f'_{nj_2}(\tau')
\ln|\sin(\pi T(\tau-\tau'))|.
\end{eqnarray}
Let us first consider the terms in the above sum which correspond to
$j_1=j_2$. They are
\begin{equation}
   -\frac{\alpha}{2}\sum\limits_na_n\int\limits_{-\beta/2}^{\beta/2}d\tau\int\limits_{-\beta/2}^{\beta/2}
   d\tau'f'_{nj}(\tau)\bar f'_{nj}(\tau')\ln|\sin(\pi T(\tau-\tau'))|.
\label{A4}
\end{equation}
Since an effective instanton width $1/\omega$ is much smaller than
the inverse temperature $\beta$ it is possible to split the
contribution (\ref{A4}) into regular and singular (in the limit $T
\to 0$) parts, respectively
\begin{equation}
  S_{\rm
    reg}=-\frac{\alpha}{2}\sum\limits_na_n\int\limits_{-\infty}^\infty
  d\tau\int\limits_{-\infty}^\infty d\tau'\frac{\partial^2
  }{\partial\tau\partial\tau'}\cos(n(\tilde\theta(\tau)-\tilde\theta(\tau')))\ln|\omega(\tau-\tau')|\simeq
\frac{4\pi\alpha r}{\kappa}
\label{A5}
\end{equation}
and
\begin{equation}
S_{\rm sing}=-2\alpha\sum\limits_na_n\sin^2\left(\frac{\pi
  n}{\kappa}\right)\ln\frac{\pi T}{\omega}=-2\alpha
K\left(\frac{2\pi}{\kappa}\right)\ln\frac{\pi T}{\omega}.
\label{A6}
\end{equation}
Consider now the remaining terms with $j_1\neq j_2$.  Provided the
distance between instantons remains much larger than their width
$\sim 1/\omega$ we find
\begin{eqnarray}
  -\alpha\sum\limits_na_n\sum\limits_{j_1<j_2}
  \Re\left[C_{j_2}\bar C_{j_1}(1-e^{\frac{2\pi i n}{\kappa}\nu_{j_2}})(1-e^{-\frac{2\pi i n}{\kappa}\nu_{j_1}})\right]
  \ln(\sin(\pi
  T(\tau_{j_2}-\tau_{j_1}))).
\label{A7}
\end{eqnarray}
Performing summation over $n$ in Eq. (\ref{A7}) and combining the result with
Eqs. (\ref{A5}) and (\ref{A6}) we arrive at Eq. (\ref{instint}).

Let us note that in the particular case $\kappa=1$ the logarithmic
interaction between instantons vanish, while in the case of
$\kappa=2$ and for large enough $r$ we get  $S_{\rm
int}[\Theta]=2\pi\alpha r(n_1+n_2)+\tilde S(\alpha )$ where the
logarithmic inter-instanton interaction $\tilde S(\alpha )$ takes
the form
\begin{eqnarray}
 \tilde S(\alpha)=-4\alpha\sum\limits_{j_1<j_2}(-1)^{j_1-j_2}
\ln\left(\frac{\sin(\pi T(\tau_{j_2}-\tau_{j_1})}{\pi T\omega^{-1}}\right).
\label{intk2}
\end{eqnarray}
Hence, the partition function for our model with $\kappa=2$ is
formally equivalent to that for the well known spin-boson model
with Ohmic dissipation \cite{Leggett,Weiss}.

\section{Perturbation theory}

Let us rewrite the partition function of our problem as
\begin{equation}
   {\cal Z}=\kappa\sum\limits_{n=0}^{\infty}\sum\limits_{\nu_1=\pm1}..\sum\limits_{\nu_n=\pm 1}
   \left(\frac{\Delta}{2}\right)^n\int\limits_0^\beta d\tau_1\int\limits_{\tau_1}^\beta d\tau_2...
   \int\limits_{\tau_{n-1}}^{\beta}d\tau_n\sum\limits_{m=-\infty}^{\infty}e^{2\pi im\phi_x-S_{\rm int}[\Theta (\tau )]}
   \delta_{\sum\limits_i \nu_i,m\kappa}.
\label{B1}
\end{equation}
With the aid of the Poisson's resummation formulae Eq. (\ref{B1})
can be transformed to
\begin{equation}
 \mathcal Z=\kappa\int\limits_{0}^{2\pi}\frac{dz}{2\pi}
 \sum\limits_{m=-\infty}^{\infty}e^{2\pi im\phi_x-im\kappa z}Z[z;\beta]=\sum\limits_{k=1}^{\kappa}Z[2\pi(\phi_x-k)/\kappa;\beta],
\end{equation}
where
\begin{equation}
 Z[z;\beta]=\sum\limits_{k=0}^{\infty}\sum\limits_{\nu_1=\pm1}..
 \sum\limits_{\nu_k=\pm 1}\left(\frac{\Delta_r}{2}\right)^n\int\limits_0^\beta d\tau_1\int\limits_{\tau_1}^\beta d\tau_2...
 \int\limits_{\tau_{n-1}}^{\beta}d\tau_n e^{iz\sum\limits_{j=1}^k\nu_j+2\alpha\sum\limits_{a<b=1}^k\nu_a\nu_bg(\varphi_{ab})W(\tau_a-\tau_b)}
\label{B3}
\end{equation}
and
\begin{equation}
W(\tau_a-\tau_b)=\ln\left[\frac{\sin[\pi
T(\tau_b-\tau_a)]}{\pi
T\omega^{-1}}\right]=-\sum\limits_{n=-\infty}^{\infty}W_ne^{2\pi
inT(\tau_a-\tau_b)}.
\end{equation}
Expanding the exponent in Eq. (\ref{B3}) one recovers the
perturbation series for $Z[z;\beta]$.  Extending the definition of
the Fourier coefficients $a_n$ to negative $n$ in such a way that
$a_{-n}=a_n$ and $\alpha_0=0$ we may write
\begin{equation}
g(\varphi_{ab})=2\sum\limits_{n=1}^ra_n \sin^2\left(\frac{\pi
n}{\kappa}\right)\cos(n\varphi_{ab})=\sum\limits_{n=-r}^ra_n
\sin^2\left(\frac{\pi n}{\kappa}\right)e^{in\varphi_{ab}}.
\end{equation}
With the aid of this equation it is easy to define all orders of
the perturbative expansion $Z=Z^{(0)}+Z^{(1)}+...$, where
\begin{equation}
Z^{(0)}[z;\beta]=e^{-2\beta\Delta_r\sin^2(z/2)}
\end{equation}
and
\begin{equation}
 Z^{(1)}[z;\beta]=-2\alpha\Delta_r^2\sum\limits_{n=-r}^ra_n\sin^2\left(z+\frac{\pi n}{\kappa}\right)\sin^2\left(\frac{\pi n}{\kappa}\right)
 \int\limits_0^\beta d\tau_1\int\limits_{\tau_1}^\beta d\tau_2Z_0[z;\tau_1]Z_0\left[z+\frac{2\pi n}{\kappa};\tau_2-\tau_1\right]
 Z_0[z;\beta-\tau_2]W(\tau_2-\tau_1).
\end{equation}
The last term can also be represented graphically by the diagram
in Fig. \ref{f4}.
\begin{figure}[t]
\includegraphics[width=2.8in]{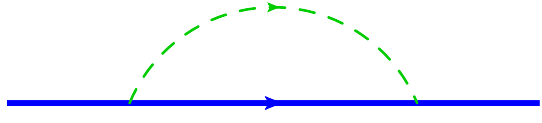}
\caption{\label{f4} (Color online) First order diagram of the
perturbation theory in $\alpha$.}
\end{figure}
Performing the Fourier transformation of the partition
function
\begin{equation}
Z[z,\beta]=\int\limits_{-\infty}^\infty\frac{d\lambda}{2\pi}e^{-i\lambda
\beta}\tilde Z[z,\lambda]
\label{B8}
\end{equation}
and introducing the self-energy $\tilde\Sigma$ as a sum of all
irreducible diagrams, we obtain
\begin{equation}
   \tilde Z[z,\lambda]=\frac{1}{-i\lambda+2\Delta_r\sin^2(z/2)-\tilde \Sigma[z;\lambda]}.
\label{B9}
\end{equation}
To the first order in the interaction one
finds
\begin{equation}
   \tilde \Sigma_1[z;\lambda]=2\alpha\Delta_r^2\sum\limits_{n=-r}^ra_n\sin^2\left(z+\frac{\pi n}{\kappa}\right)
   \sin^2\left(\frac{\pi n}{\kappa}\right)\sum\limits_{k=-\infty}^\infty\frac{W_k}{-i\lambda-2\pi i kT+2\Delta_r\sin^2(\frac{z}{2}
   +\frac{\pi n}{\kappa})},
\end{equation}
where  $W_0=\ln[2\pi T/\omega]$ and  $W_k=1/2|k|$ for $k\neq0 $.
Performing summation over $k$, we get
\begin{eqnarray}
\tilde\Sigma_1[z;\lambda]=2\alpha\Delta_r^2\sum\limits_{n=-r}^ra_n\sin^2\left(z+\frac{\pi
n}{\kappa}\right)\sin^2\left(\frac{\pi n}{\kappa}\right)
   \qquad\qquad\qquad\qquad\qquad\qquad\qquad\nonumber\\ \times
\frac{\ln\frac{2\omega}{\pi
T}+\gamma+\frac12\psi\left(1+\frac{\lambda+2i\Delta_r\sin^2(\frac{z}{2}+\frac{\pi
n}{\kappa})}{2\pi
T}\right)+\frac12\psi\left(1-\frac{\lambda+2i\Delta_r\sin^2(\frac{z}{2}+\frac{\pi
n}{\kappa})}{2\pi
T}\right)}{-i\lambda+2\Delta_r\sin^2(\frac{z}{2}+\frac{\pi
n}{\kappa})},
\end{eqnarray}
where $\psi$ is the digamma function and $\gamma \simeq 0.577$ is
the Euler constant. Substituting this expression into Eq.
(\ref{B9}), we observe that $\tilde Z[z,\lambda]$ has a pole at
$\lambda_p=-2i\Delta_r\sin^2(z/2)+\delta\lambda_p$, where
\begin{eqnarray}
\delta\lambda_p=i\alpha\Delta_r\sum\limits_{n=-r}^ra_n\sin\left(z+\frac{\pi
n}{\kappa}\right)\sin\left(\frac{\pi
n}{\kappa}\right)\times\qquad\qquad\qquad\qquad\qquad\qquad\qquad\qquad\qquad\qquad\qquad\qquad\nonumber\\
\times \left[\ln\frac{2\pi
T}{\omega}+\gamma+\frac12\psi\left(1+\frac{i\Delta_r}{\pi
T}\sin\left(z+\frac{\pi n}{\kappa}\right)\sin\left(\frac{\pi
n}{\kappa}\right)\right)+\frac12\psi\left(1-\frac{i\Delta_r}{\pi
T}\sin\left(z+\frac{\pi n}{\kappa}\right)\sin\left(\frac{\pi
n}{\kappa}\right)\right)\right].
\end{eqnarray}
Combining the above results with Eq. (\ref{B8}) one arrives at the
final perturbative in the interaction expression for the partition
function. Of particular interest is the low temperature limit
$T\ll\Delta_r/\kappa^2$. In this case one has
$\psi(1+z)\approx\ln(z)+O(1/z)$ and the free energy reduces to
\begin{eqnarray}
   -T \ln\mathcal Z=2\Delta_r\sin^2\left(\frac{\pi\phi_x}{\kappa}\right)-\alpha\Delta_r\sum\limits_{n=-r}^ra_n
   \sin\left(\frac{\pi n}{\kappa}\right)\sin\left(\frac{2\pi\phi_x+\pi n}{\kappa}\right)\qquad\qquad\nonumber\\
   \times\left[\ln\frac{2\Delta e^\gamma}{\omega}+
   \ln\left|\sin\left(\frac{\pi n}{\kappa}\right)\sin\left(\frac{2\pi\phi_x+\pi
   n}{\kappa}\right)\right|\right].
\label{B13}
\end{eqnarray}
It is straightforward to observe that the first logarithmic term
in the square brackets simply yields the renormalization
$\Delta_r\rightarrow \Delta_r\left(1+2\alpha
K(2\pi/\kappa)\ln\frac{2\Delta_r e^\gamma}{\omega}\right)$ and
does not alter the dependence of PC on the flux $\phi_x$. Hence,
this perturbative contribution can be absorbed in the first term
in (\ref{B13}) simply by substituting $\Delta_R$ instead of
$\Delta_r$, where $\Delta_R$ is defined in Eq. (\ref{renlt}).

The second logarithmic term in the square brackets in (\ref{B13})
contains an additional flux dependence which turns singular at
$\phi_x$ close to the half-integer numbers. Clearly, this
perturbative contribution cannot be just reduced to the
renormalization (\ref{renlt}). As a result, we obtain
\begin{eqnarray}
   -T \ln\mathcal Z=2\Delta_R\sin^2\left(\frac{\pi\phi_x}{\kappa}\right)-\alpha\Delta_r\sum\limits_{n=-r}^ra_n
   \sin\left(\frac{\pi n}{\kappa}\right)\sin\left(\frac{2\pi\phi_x+\pi n}{\kappa}\right)
   \ln\left|\sin\left(\frac{\pi n}{\kappa}\right)\sin\left(\frac{2\pi\phi_x+\pi
   n}{\kappa}\right)\right|.
\label{B14}
\end{eqnarray}
Employing this expression we arrive at Eq. (\ref{PCT0}).
\end{widetext}

\end{document}